\documentstyle[11pt]{article}

\def\be{\begin{equation}}
\def\ee{\end{equation}}
\def\ba{\begin{eqnarray}}
\def\ea{\end{eqnarray}}
\def\bann{\begin{eqnarray*}}
\def\eann{\end{eqnarray*}}
\def\mref#1{Eq.~(\ref{eq:#1})}
\def\nref#1{(\ref{eq:#1})}
\def\oref#1{\ref{eq:#1}}

\def\mlab#1{\label{eq:#1}}
\def\nn{\nonumber}
\def\Comment#1{}

\def\F{{\cal F}}

\def\l{\left}
\def\r{\right}

\def\half{\mbox{\small{$\frac{1}{2}$}}}

\def\F21{\mbox{}_2F_1}

\begin{document}

\baselineskip=20pt

\hfill RUGTh-980204

\begin{center}
{\LARGE \bf
QCD in the Infrared with Exact Angular Integrations\\
}
\vspace{6mm}

{\bf D. Atkinson\footnote{atkinson@phys.rug.nl} and
J.C.R. Bloch\footnote{bloch@phys.rug.nl}\\
Institute for Theoretical Physics\\
University of Groningen\\
Nijenborgh~4\\
NL-9747~AG~~Groningen\\
The Netherlands
}
\end{center}

{\bf Abstract}
\begin{center}
{\parbox{120mm}{\small In a previous paper we have shown that in quantum
chromodynamics the gluon propagator vanishes in the infrared limit, while
the ghost propagator is more singular than a simple pole.  
These results were obtained after angular averaging, but in 
the current paper we go beyond this approximation 
and perform an exact calculation of the angular
integrals. The powers of the infrared behaviour of the
propagators are changed substantially. We find the very intriguing result
that the gluon propagator vanishes in the infrared 
exactly like $p^2$, whilst the ghost
propagator is exactly as singular as $1/p^4$. We also
find that the value of the infrared fixed point of the QCD coupling
is much decreased from the y-max estimate: 
it is now equal to $4\pi/3$.}}
\end{center}

\vspace{0.5cm}

Following a recent study by von Smekal et al.\cite{Smekal}, we
analyzed in Ref.\cite{AB97} the coupled Dyson-Schwinger equations
for the gluon and ghost form factors $F$ and $G$. The approximations
were two-fold: firstly the vertices were taken bare, and secondly angular 
averaging was introduced (the so-called
y-max approximation). Deferring to later work an improvement
of the vertices, in this paper we seek to remove the deficiency of the
y-max approximation. On the one hand the results might be regarded
simply as quantitative adjustments to the y-max calculations; but on the
other hand they are far from negligible. The numerical value of the
infrared fixed point is reduced by a factor of almost three; and the
finding that the gluon propagator has a simple zero, while the ghost
propagator has a double pole, might perhaps be deemed a qualitatively
new result.
 
As an improvement on the y-max approximation used in Ref.\cite{AB97}, we now
solve the coupled integral equations for the gluon and ghost propagators
with an exact treatment of the angular integrals. 
Although the y-max approximation is good in the ultraviolet region, where
the form factors run logarithmically, we will see that 
it is a crude approximation in the
infrared region, where the form factors exhibit power behaviour. The y-max
approximation apparently leads to a difficulty in the ghost equation, as
the asymmetry in the treatment of the gluon and ghost momenta in the loop
gives an ambiguous result. In our previous study we used the ghost momentum
for the radial integration and the gluon momentum for the angular
integration. However, if we exchange these momenta we no longer find a
consistent solution to the integral equation. We will show later
why the first choice is a better approximation than the latter, thus
motivating the choice made in Ref.\cite{AB97}. Furthermore we will show
that both choices are equivalent if we treat the angular integrals exactly,
although this equivalence is by no means trivial.

With the radial integration over the ghost
momentum, the ghost form factor satisfies
\be
\frac{1}{G(x)} = \tilde{Z}_3 - \frac{6\lambda}{\pi}\int_0^{\Lambda^2} dy\,
y G(y) \int_0^\pi d\theta \, \sin^4\theta\,\frac{F(z)}{z^2} \,,
\mlab{Gh1}
\ee
whereas with the radial integration over the gluon momentum, we have instead 
\be
\frac{1}{G(x)} = \tilde{Z}_3 - \frac{6\lambda}{\pi}\int_0^{\Lambda^2} dy\,
F(y) \int_0^\pi d\theta \, \sin^4\theta\,\frac{G(z)}{z} \,,
\mlab{Gh2}
\ee
where $\lambda=g^2/16\pi^2$, $g$ being the strong coupling constant and 
where $z=x+y-2\sqrt{xy}\cos\theta$.
Here we have used the fact that $\tilde Z_1$, the ghost-gluon vertex
renormalization constant, is equal to unity in the Landau gauge. 

We will show that
\be
F(x) = A x^{2\kappa} \hspace{2cm} G(x) = B x^{-\kappa}
\mlab{sol}
\ee
are solutions of \mref{Gh1} and \mref{Gh2}, yielding identical conditions on
$\lambda A B^2$, but that this equivalence {\em does not hold} 
in the y-max approximation. 

Substituting \mref{sol} in \mref{Gh1} and evaluating the angular
integral in terms of the hypergeometric function, $\F21$, 
we obtain
\bann
\frac{x^{\kappa}}{B} &=& \tilde{Z}_3 
- \frac{9\lambda AB}{4}\int_0^{\Lambda^2} dy\,
y^{-\kappa+1} y_>^{2\kappa-2} \,\F21\l(-2\kappa+2,-2\kappa;3;\frac{y_<}{y_>}\r) \\
&=& \tilde{Z}_3 - \frac{9\lambda AB}{4}\Bigg\{
x^{2\kappa-2} \int_0^x dy\, y^{-\kappa+1} 
\,\F21\l(-2\kappa+2,-2\kappa;3;\frac{y}{x}\r) \\
&& + \int_x^{\Lambda^2} dy\, y^{\kappa-1} 
\,\F21\l(-2\kappa+2,-2\kappa;3;\frac{x}{y}\r) \Bigg\}\,.
\eann
Set $t=y/x$ in the infrared integration and $t=x/y$ in the ultraviolet
integration and take $\Lambda^2 \to \infty$: 
\be
\frac{x^{\kappa}}{B} =  - \frac{9\lambda AB}{4} x^{\kappa}
\int_0^1 dt\, \l( t^{-\kappa+1} + t^{-\kappa-1} \r)
\F21\l(-2\kappa+2,-2\kappa;3;t\r) \,.
\mlab{Gh1a}
\ee
Note that we have dropped $\tilde{Z}_3 $. This is in accordance with 
standard regularization procedure: the integral in \mref{Gh1} is
convergent if Re $\kappa <0$, whereas a subtraction is necessary if 
$0\le \kappa <1$. By identifying $\tilde{Z}_3 $ with this subtraction
constant, we ensure that $G(x)$ is defined by analytic continuation in
$\kappa$ beyond Re $\kappa =0$. This continuation is made explicit in terms
of the generalized hypergeometric function 
$\mbox{}_3F_2$
(see \mref{intF}). 
After matching the coefficients of $x^\kappa$
we obtain 
\ba
\mlab{Gh1b}
\frac{1}{\lambda A B^2} &=& 
- \frac{9}{4} \l[ \frac{1}{2-\kappa}
\;\mbox{}_3F_2\l(-2\kappa+2,-2\kappa,2-\kappa;3,3-\kappa;1\r) \r. \\
&& \l. \hspace{8mm} - \frac{1}{\kappa}
\;\mbox{}_3F_2\l(-2\kappa+2,-2\kappa,-\kappa;3,1-\kappa;1\r) \r] \,. \nn
\ea

The y-max approximation that we made in \cite{AB97} amounts to replacing
$\kappa$ by zero in the hypergeometric function in \mref{Gh1a}. The result 
was
\be
\frac{1}{\lambda A B^2} = 
- \frac{9}{4} \l[ \frac{1}{2-\kappa} - \frac{1}{\kappa}\r] \,. 
\mlab{GFymax}
\ee

Next we make a similar analysis of  \mref{Gh2}: 
\be
\frac{x^{\kappa}}{B} = \tilde{Z}_3 
- \frac{9\lambda AB}{4}\int_0^{\Lambda^2} dy\,
y^{2\kappa} y_>^{-\kappa-1} \,\F21\l(\kappa+1,\kappa-1;3;\frac{y_<}{y_>}\r) \,.
\ee
After 
defining $t=y/x$ in the infrared integration and $t=x/y$ in the ultraviolet
integration, taking $\Lambda^2 \to \infty$ and 
eliminating $\tilde{Z}_3$ as before, we obtain 
\be
\frac{x^{\kappa}}{B} = - \frac{9\lambda AB}{4} x^{\kappa}
\int_0^1 dt\, \l( t^{2\kappa} + t^{-\kappa-1} \r)
\F21\l(\kappa+1,\kappa-1;3;t\r) \,.
\mlab{Gh2a}
\ee

Using \mref{intF} in \mref{Gh2a} and equating the coefficients of $x^\kappa$
we find
\ba
\mlab{Gh2b}
\frac{1}{\lambda A B^2} &=& 
- \frac{9}{4} \l[ \frac{1}{2\kappa+1} 
\;\mbox{}_3F_2\l(\kappa+1,\kappa-1,2\kappa+1;3,2\kappa+2;1\r) \r. \\
&& \hspace{9mm} \l. - \frac{1}{\kappa}
\;\mbox{}_3F_2\l(\kappa+1,\kappa-1,-\kappa;3,1-\kappa;1\r) \r] \nn \,.
\ea

The y-max approximation of this expression is once more obtained by replacing
$\kappa$ by zero in the hypergeometric function in \mref{Gh2a}, 
yielding 
\be
\frac{1}{\lambda A B^2} = 
- \frac{9}{4} \l[ \frac{1}{2\kappa+1}  -\frac{1}{3(2\kappa+2)}
- \frac{1}{\kappa} + \frac{1}{3(\kappa-1)} \r] \, .
\mlab{FGymax}
\ee
This approximation is quite different from that of \mref{GFymax},
illustrating the precariousness of the y-max approximation.  

However, the exact forms \mref{Gh1b} and \mref{Gh2b} are identical, as
we verified by using Mathematica to check the numerical
equivalence over a domain of $\kappa$ values. 

We now consider the gluon equation, and will show that it too is solved
by a pure
power behaviour if we keep only the ghost loop. 
Even in the presence of the gluon loop, when the power behaviour is
no longer an exact solution of the equation, 
it will still represent the correct
leading order infrared asymptotic behaviour. Omitting then the gluon loop,
we write the 
equation for the gluon form factor as follows:
\be
\mlab{Gl}
\frac{1}{F^(x)} = Z_3 + \frac{2\lambda}{\pi}
\int_0^{\Lambda^2}\frac{dy}{x}G(y)\int_0^\pi d\theta 
\sin^2\theta \, M(x,y,z)G(z) \,,
\ee
where 
\[
M(x,y,z) = 
\left(\frac{x+y}{2}-\frac{y^2}{x}\right)\frac{1}{z} 
+\frac{1}{2}+\frac{2y}{x}-\frac{z}{x}
\, .
\]

We now substitute the solution \nref{sol} into \mref{Gl}, obtaining
\[
\frac{x^{-2\kappa}}{A} = Z_3 + \frac{2\lambda B^2}{\pi}
\int_0^{\Lambda^2}\frac{dy}{x} y^{-\kappa} \int_0^\pi d\theta 
\sin^2\theta \, \l[ \left(\frac{x+y}{2}-\frac{y^2}{x}\right)z^{-\kappa-1} 
+ \l(\frac{1}{2}+\frac{2y}{x}\r)z^{-\kappa} - \frac{z^{-\kappa+1}}{x}\r] \,.
\]

We evaluate the angular integral with the help of \mref{angint}:
\bann
\frac{x^{-2\kappa}}{A} &=& Z_3 + \lambda B^2
\int_0^{\Lambda^2}\frac{dy}{x} y^{-\kappa}
\l[ \left(\frac{x+y}{2}-\frac{y^2}{x}\right) 
y_>^{-\kappa-1} \,\F21\l(\kappa+1,\kappa;2;\frac{y_<}{y_>}\r) \r.\\
&& \l. + \l(\frac{1}{2}+\frac{2y}{x}\r) 
y_>^{-\kappa} \,\F21\l(\kappa,\kappa-1;2;\frac{y_<}{y_>}\r)
- \frac{y_>^{-\kappa+1}}{x}\,\F21\l(\kappa-1,\kappa-2;2;\frac{y_<}{y_>}\r)\r] \,.
\eann

We split the integration region and define 
$t=y/x$ in the infrared integration and $t=x/y$ in the ultraviolet
integration, as before, and finally implicitly implement the 
analytic regularization, taking $\Lambda^2 \to \infty$. 
After some rearrangement we obtain 
\ba
\mlab{Gla}
\lefteqn{\frac{x^{-2\kappa}}{A} =  \lambda B^2 x^{-2\kappa}
\int_0^1 dt} \\
&& \l\{
\left[\half \l(t^{-\kappa}+t^{-\kappa+1}+t^{2\kappa-1}+t^{2\kappa-2}\r) 
- \l(t^{-\kappa+2} + t^{2\kappa-3}\r)\right] 
\F21\l(\kappa+1,\kappa;2;t\r) \r.\nn\\
&& \l. + \l[\half \l(t^{-\kappa}+t^{2\kappa-2}\r)
+ 2 \l(t^{-\kappa+1}+t^{2\kappa-3}\r)\r] \F21\l(\kappa,\kappa-1;2;t\r) \r. 
\nn\\
&& \l. - \l[t^{-\kappa}+t^{2\kappa-3}\r] \F21\l(\kappa-1,\kappa-2;2;t\r)\r\} 
\,.\nn
\ea

Using \mref{intF} in \mref{Gla} and equating coefficients of $x^\kappa$, 
we find
\ba
\mlab{Glb}
\lefteqn{\frac{1}{\lambda A B^2} = } \\ 
&& \frac{1}{2(1-\kappa)}
\;\mbox{}_3F_2\l(\kappa+1,\kappa,1-\kappa;2,2-\kappa;1\r)
+ \frac{1}{2(2-\kappa)}
\;\mbox{}_3F_2\l(\kappa+1,\kappa,2-\kappa;2,3-\kappa;1\r)  \nn\\
&& + \frac{1}{4\kappa}
\;\mbox{}_3F_2\l(\kappa+1,\kappa,2\kappa;2,2\kappa+1;1\r)
+ \frac{1}{2(2\kappa-1)}
\;\mbox{}_3F_2\l(\kappa+1,\kappa,2\kappa-1;2,2\kappa;1\r)  \nn\\
&& - \frac{1}{3-\kappa}
\;\mbox{}_3F_2\l(\kappa+1,\kappa,3-\kappa;2,4-\kappa;1\r)
- \frac{1}{2(\kappa-1)}
\;\mbox{}_3F_2\l(\kappa+1,\kappa,2\kappa-2;2,2\kappa-1;1\r)  \nn\\
&& + \frac{1}{2(1-\kappa)}
\;\mbox{}_3F_2\l(\kappa,\kappa-1,1-\kappa;2,2-\kappa;1\r)
+ \frac{1}{2(2\kappa-1)}
\;\mbox{}_3F_2\l(\kappa,\kappa-1,2\kappa-1;2,2\kappa;1\r)  \nn\\
&& + \frac{2}{2-\kappa}
\;\mbox{}_3F_2\l(\kappa,\kappa-1,2-\kappa;2,3-\kappa;1\r)
+ \frac{1}{\kappa-1}
\;\mbox{}_3F_2\l(\kappa,\kappa-1,2\kappa-2;2,2\kappa-1;1\r)  \nn\\
&& - \frac{1}{1-\kappa}
\;\mbox{}_3F_2\l(\kappa-1,\kappa-2,1-\kappa;2,2-\kappa;1\r)
- \frac{1}{2(\kappa-1)}
\;\mbox{}_3F_2\l(\kappa-1,\kappa-2,2\kappa-2;2,2\kappa-1;1\r) \,. \nn
\ea
Some of these generalized hypergeometric functions are in fact infinite for
certain values of $\kappa$; they should be understood by replacing the
final argument `1' in each 
generalized hypergeometric function by $\tau$, and then by taking the
limit $\tau\rightarrow 1$. Since this amounts to 
replacing the upper integration limit in \mref{Gla} by $\tau$, 
the procedure is clearly legitimate. 

Finally, to demonstrate that \mref{sol} really is a solution of the coupled
integral equations, for a particular value of $\kappa$, 
we equate the right-hand side of \mref{Gh1b} or
\mref{Gh2b} with that of \mref{Glb} and solve for $\kappa$. We 
used the routine FindRoot of Mathematica and discovered that, in the limit 
\[
\kappa \to 1 \, ,
\mlab{kappato1}
\]
both expressions become equal. This limit has been checked 
analytically in App.~\ref{Sect:limab2}. We find 
\[
\lambda A B^2 = \frac{1}{3}
\]
which means that the infrared fixed point for the gauge invariant running 
coupling, $\alpha(p^2) = 4\pi \lambda \tilde{Z}_1^2 F(p^2) G^2(p^2)$, 
has the value 
\[
\alpha(0) = \frac{4\pi}{3} \approx 4.19
\]
which is quite different from the the value we found using the y-max
approximation. There we had $\alpha(0) \approx 11.47$, nearly three
times the exact result, corresponding to $\kappa\approx 0.77$.

Our findings for the propagators in the infrared region can be summarized 
by the following formulas:

gluon propagator in Landau gauge
\be
D^{ab}_{\mu\nu}(p) \sim - \delta^{ab}
\l[g_{\mu\nu} - \frac{p_\mu p_\nu}{p^2}\r] p^2 \,,
\ee

ghost propagator in Landau gauge
\be
G^{ab}(p) \sim -\delta^{ab} \frac{1}{p^4} \,.
\ee

\appendix

\section {Angular and Radial Integrations}

Throughout we used the angular integration formula
\be
\int_0^\pi d\theta \, \sin^{2r}\theta \, z^n =
B\l(r+\frac{1}{2},\frac{1}{2}\r) \, y_>^n \,
\,\F21\l(-n,-n-r;r+1;\frac{y_<}{y_>}\r) \,.
\mlab{angint}
\ee

The radial integrations of hypergeometric functions are evaluated as
follows: 
\ba
\int_0^1 dt\, t^\nu \,\F21\l(a,b;c;t\r)
&=& \sum_{n=0}^{\infty} \frac{(a)_n \, (b)_n}{(c)_n \, n!} \int_0^1 dt\, 
t^{\nu+n} \nn\\
&=& \sum_{n=0}^{\infty} \frac{(a)_n (b)_n}{(c)_n \, n!} \frac{1}{\nu+n+1} \nn\\
&=& \frac{1}{\nu+1} \, \mbox{}_3F_2\l(a,b,\nu+1;c,\nu+2;1\r) \,.
\mlab{intF}
\ea

\section {Limit of $\lambda AB^2$ for $\kappa\to 1$}
\label{Sect:limab2}

We check analytically that the limit of $\lambda AB^2$ for $\kappa\to 1$ is
identical for Eqs.~(\oref{Glb}, \oref{Gh1b}, \oref{Gh2b}). 
Starting from
\mref{Gh1b} we find
\bann
\frac{1}{\lambda A B^2} &=&
- \frac{9}{4} \bigg\{
\;\mbox{}_3F_2\l(0,-2,1;3,2;1\r)
- \lim_{\kappa\to 1} \;\mbox{}_3F_2\l(2-2\kappa,-2,-1;3,1-\kappa;1\r) 
\bigg\}
\\
&=& - \frac{9}{4}  \l[ 
1 - \l( 1 + \frac{4}{3} \r) \r] = 3
\eann

and from \mref{Gh2b}
\bann
\frac{1}{\lambda A B^2} &=&
- \frac{9}{4} \bigg\{
 \frac{1}{3} 
\;\mbox{}_3F_2\l(2,0,3;3,4;1\r)
- \lim_{\kappa\to 1} \;\mbox{}_3F_2\l(2,\kappa-1,-1;3,1-\kappa;1\r) 
\bigg\}
 \\
&=& - \frac{9}{4} \l[ \frac{1}{3} - \l(1 + \frac{2}{3} \r) \r] = 3 \,.
\eann

We make an analoguous check for the gluon equation. 
To circumvent convergence problems, we replace the argument of the 
generalized hypergeometric functions by $\tau$, and take the limit
$\tau\to 1$ at the end.
From  \mref{Glb}, 
\bann
\lefteqn{\frac{1}{\lambda A B^2} =} \\
&& \lim_{\tau\to 1}\lim_{\kappa\to 1}\bigg\{
\frac{1}{2(1-\kappa)} \;\mbox{}_3F_2\l(2,1,1-\kappa;2,1;\tau\r)
+ \mbox{}_3F_2\l(2,1,1;2,2;\tau\r) 
- \frac{1}{4} \;\mbox{}_3F_2\l(2,1,2;2,3;\tau\r)
\\
&& 
- \frac{1}{2(\kappa-1)} \;\mbox{}_3F_2\l(2,1,2\kappa-2;2,1;\tau\r) 
+ \frac{1}{2(1-\kappa)} \;\mbox{}_3F_2\l(1,\kappa-1,1-\kappa;2,1;\tau\r)
\\
&& 
+ \frac{5}{2} \;\mbox{}_3F_2\l(1,0,1;2,2;\tau\r) 
+ \frac{1}{\kappa-1} \;\mbox{}_3F_2\l(1,\kappa-1,2\kappa-2;2,1;\tau\r) 
\\
&& - \frac{1}{1-\kappa} \;\mbox{}_3F_2\l(\kappa-1,-1,1-\kappa;2,1;\tau\r)
- \frac{1}{2(\kappa-1)} \;\mbox{}_3F_2\l(\kappa-1,-1,2\kappa-2;2,1;\tau\r) 
\bigg\} \\[5pt]
&=&\lim_{\tau\to 1}\lim_{\kappa\to 1}\bigg\{
\frac{1}{2(1-\kappa)} \sum_{n=1}^\infty \frac{(1-\kappa)_n
+(2\kappa-2)_n}{n!}\tau^n
+ \;\mbox{}_2F_1\l(1,1;2;\tau\r) 
- \frac{1}{4}\;\mbox{}_2F_1\l(1,2;3;\tau\r)
\bigg\}  + \frac{5}{2} \\[5pt]
&=&  \lim_{\tau\to 1}\bigg\{
\sum_{n=1}^\infty \l( -\frac{1}{2n} + \frac{1}{n+1} - 
\frac{1}{2(2+n)} \r) \tau^n
\bigg\} + \frac{13}{4} \\[5pt]
&=&  
3 + \lim_{\tau\to 1}
\bigg\{\sum_{n=3}^\infty \l( -\frac{1}{2n} + \frac{1}{n} - 
\frac{1}{2n} \r)\tau^n \bigg\}  \\[5pt]
&=&  3 \,.
\eann

{\bf Acknowledgments}

The authors thank the CSSM/NITP hosted by the University of Adelaide for their
hospitality during the Workshop on Nonperturbative Methods in Quantum
Field Theory, during which this paper was completed. J.C.R.~Bloch was
supported by F.O.M. ({\it Stichting voor Fundamenteel Onderzoek der
Materie}).

\raggedright

\end{document}